\documentclass[aps,prl,showpacs,twocolumn,floatfix]{revtex4}%
\usepackage{amssymb}
\usepackage{graphics}
\usepackage{amsmath}
\usepackage{graphicx}
\usepackage{amsfonts}

\begin{document}
\title{Side-channel-free quantum key distribution}
\author{Samuel L. Braunstein}
\affiliation{Computer Science, University of York, York YO10 5GH,
United Kingdom}
\author{Stefano Pirandola}
\affiliation{Computer Science, University of York, York YO10 5GH,
United Kingdom}
\date{\today }

\begin{abstract}
Quantum key distribution (QKD) offers the promise of absolutely secure
communications. However, proofs of absolute security often assume
perfect implementation from theory to experiment. Thus, existing systems
may be prone to insidious side-channel attacks that rely on flaws in
experimental implementation.
Here we replace all real channels with virtual channels in a QKD
protocol, making the relevant detectors and settings inside
private spaces inaccessible while simultaneously acting as a
Hilbert space filter to eliminate side-channel attacks. By using a
quantum memory we find that we are able to bound the secret-key
rate below by the entanglement-distillation rate computed over the
distributed states.
\end{abstract}


\maketitle

In 1982 Richard Feynman conjectured the use of quantum systems as
a technological platform for solving difficult calculations in
physics. Eventually this insight lead to the field of quantum
information processing. As part of the field's growth, it has
partly diverged into the two main application domains: computation
and communications, though much fundamental and technical overlap
still exists. Interestingly, the key application that has started
to mature and is now commercially available is quantum
cryptography, or more precisely quantum key distribution (QKD)
which has quickly moved from the purely theoretical
\cite{BB84,Ekert,Hillery,Cerf} to a practical technology
\cite{Grangier,Weed,Scarani,Net,GQI}.

How can we explain the impressive industrial uptake of quantum
cryptography and its ultimate aim to take over classical systems?
The answer lies in the claim of ``absolute security''
\cite{Gis02}. Unfortunately, while the idea is very compelling,
subtle details in implementation may introduce flaws that could,
potentially, be open to attack. Specifically, attacks from so
called ``side channels'' represent one of the most elusive threats
in practical quantum cryptography, because a system could be
vulnerable to side-channel attacks even if it is unbreakable in
theory \cite{Lut09,SCA}. In fact, the recent approach of
``device-independent QKD'' \cite{DIQKD} makes important advances
in handling imperfect implementations, and can even be made by
untrusted parties, but does not directly address all possible
side-channel attacks, where, for example, detectors may directly
receive external probing aimed at seeding or gleaning their
readout.

In principle side-channel attacks affect both classical and quantum
cryptography, but could be especially devastating for quantum
cryptography, precisely because of the proclaimed absolute security
``guarantee''. The threat from such attacks has been demonstrated
in both lab and installed field settings \cite{SCA}.
%
%
Thus, while practical QKD systems have
been fighting a trade-off between distance and key generation rate,
they are still facing the fundamental problem of guaranteed security,
choosing to rely on theoretical promises of absolute security without
having any way of authenticating them in practice.

\begin{figure}[hptb]
\vspace{0.0cm}
\par
\begin{center}
\includegraphics[width=0.47\textwidth]{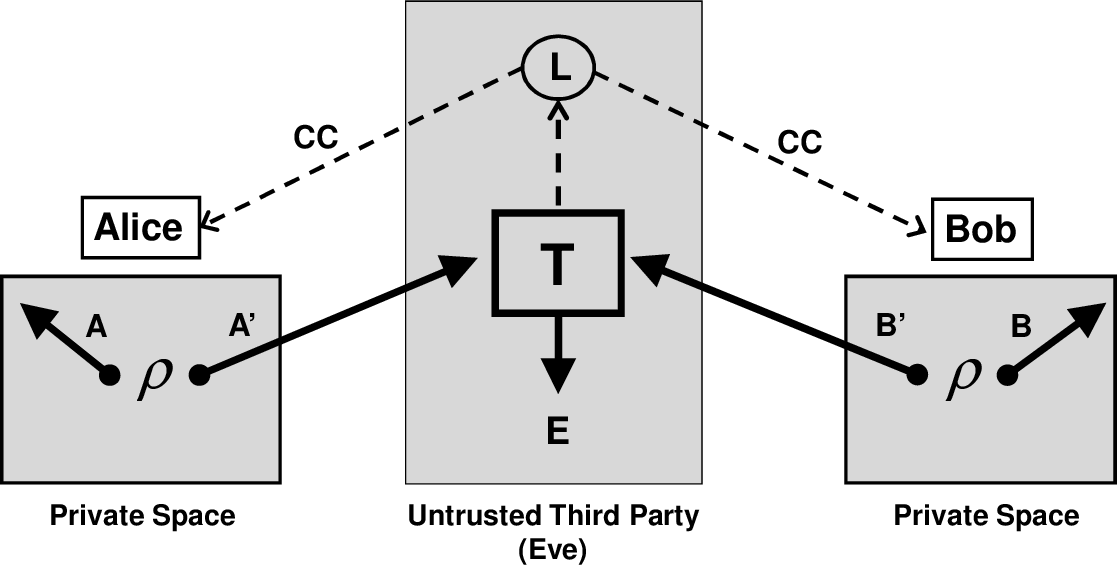}
\end{center}
\par
\vspace{-0.5cm}\caption{Private space to private space. The UTP\ acts as a
correlator.}%
\label{GEN}%
\end{figure}

\vskip 0.0truein
\noindent
{\bf Private spaces: general model}

Let us consider the scenario of Fig.~\ref{GEN}. Two authenticated
parties, Alice and Bob, control two private spaces, $\mathcal{A}$
and $\mathcal{B}$, respectively. Conventionally, these spaces are
assumed completely inaccessible from the outside, i.e., no
illegitimate system may enter $\mathcal{A}$ or $\mathcal{B}$. For
this reason every kind of side-channel attack upon the private
spaces is assumed excluded. In practice, however, any port can
allow a side-channel to enter possibly probing any detector,
state-generation or detector settings. To prevent or overcome such
attacks, the QKD system must effectively isolate its private
spaces: the private space must not be directly involved in either
state preparation (for sending) or detection (of incoming states).
To overcome such probing side-channel attacks, we propose
performing state-generation by collapse of a bipartite entangled
state, so that any probe from outside is perfectly isolated from
the state-generation ``machinery'' (see Supplementary Material for
an extended discussion). Thus, in a manner akin to teleportation,
we replace all real channels with virtual channels. This allows us
to physically (and ``topologically'') separate all detectors and
settings within the private space from external probing, while
also acting as a Hilbert space filter \cite{LoChau} against any
side channel.

Within its own private space, each party (Alice or Bob) has a
bipartite state $\rho$\ which entangles two systems: $\{A,A^{\prime}\}$
for Alice, and $\{B,B^{\prime}\}$ for Bob. Systems $\{A,B\}$ are kept
within the private spaces, while systems $\{A^{\prime },B^{\prime}\}$
are sent to an untrusted third party (UTP), whose task is to perform a
quantum measurement and communicate the corresponding result. This
untrusted LOCC then allows the creation of correlations between
the private systems $\{A,B\}$ that Alice and Bob can exploit to generate
a secret-key. In its simplest form an ideal side-channel free QKD scheme
reduces to an entanglement swapping setup \cite{Biham}, with the
dual teleportation channel acting as an ideal Hilbert space filter.
What is unique about our protocol is the ability to completely protect
private space settings and detectors from probing side-channel attacks.

In the worst case scenario, the UTP must be identified with Eve herself,
whose aim is to eavesdrop the key, or even prevent Alice and Bob from
generating the key (i.e., a denial of service). In the most general
case, Eve applies a quantum instrument
$\mathbf{T}=\{T_{l}\}_{l=1}^{l_{\max}}$ to the incoming systems
$\{A^{\prime},B^{\prime}\}$. This is a quantum operation with both
classical and quantum outputs. For each classical outcome $l$, there
is a corresponding completely positive (CP) map $T_{l}$ applied to
the systems $\{A^{\prime },B^{\prime}\}$ \cite{CPTP}. This means that
the global input state $\rho_{AA^{\prime}}\otimes\rho_{BB^{\prime}}$
is transformed into the conditional output state
\begin{equation}
\rho_{ABE}(l)\equiv
\frac{1}{p(l)}(I_{A}\otimes I_{B}\otimes T_{l})(\rho_{AA^{\prime}}
\otimes\rho_{BB^{\prime}}), \label{condSTATE}
\end{equation}
where $E$\ represents an output quantum system in the hands of Eve,
while $I_{A}\otimes I_{B}$ is the identity channel acting on the
private systems $\{A,B\}$. Cleary each outcome $l$ will be found
with some probability $p(l)$, depending both on $T_{l}$ and the input
state. As a consequence the classical output of $\mathbf{T}$ can be
simply represented by the stochastic variable $L\equiv\{l,p(l)\}$.
The quantum output of $\mathbf{T}$ is represented by the system
$E$ which is correlated with the private systems $\{A,B\}$ via the
conditional state $\rho_{ABE|L}$ specified by Eq.~(\ref{condSTATE}).
$E$ is the system that Eve will use for eavesdropping. For instance,
most generally Eve can store all the output systems $E$ (generated in
many independent rounds of the protocol) into a big quantum memory.
Then, she can detect the whole memory using an optimal quantum
measurement (corresponding to a collective attack).

According to the agreed protocol, the UTP must send a classical
communication (CC) to both Alice and Bob in order to
``activate'' the correlations. Here, Eve has another weapon in
her hands, i.e., tampering with the classical outcomes. In order to
decrease the correlations between the honest parties, Eve may process
the output stochastic variable $L$ via a classical channel
\begin{equation}
p(l^{\prime}|l):L\rightarrow L^{\prime},
\end{equation}
and then communicate the fake variable
$L^{\prime}=\{l^{\prime},p(l^{\prime})\}$ to Alice and Bob, where
\begin{equation}
p(l^{\prime})=\sum_{l}p(l^{\prime},l),
\qquad p(l^{\prime},l)=p(l^{\prime}|l)p(l).
\end{equation}
This process projects the private systems $\{A,B\}$\ onto the conditional
state
\begin{equation}
\rho_{AB|L^{\prime}}=\mathrm{Tr}_{E}\left(  \rho_{ABE|L^{\prime}}\right)
,\label{ABstate}
\end{equation}
where
\begin{equation}
\rho_{ABE}(l^{\prime}) \equiv\frac{1}{p(l^{\prime})}
\!\sum_{l}p(l^{\prime},l)
\rho_{ABE}(l)  =\!\sum_{l}p(l|l^{\prime})\rho_{ABE}(l).
\end{equation}
Notice that, if $L^{\prime}$ is completely unrelated to $L$, then Eve
realizes a denial of service, being the communication of the fake
variable equivalent to tracing over systems $\{A^{\prime},B^{\prime}\}$.
In other words, for $p(l^{\prime},l)=p(l^{\prime})p(l)$, we have
$\rho_{AB|L^{\prime} }=\rho_{A}\otimes\rho_{B}$, where
$\rho_{A}\equiv \mathrm{Tr}_{A^{\prime}}\left(\rho_{AA^{\prime}}\right)$
and
$\rho_{B}\equiv\mathrm{Tr}_{B^{\prime}}\left(\rho_{BB^{\prime}}\right)$.

\vskip 0.1truein
\noindent
{\bf Secret-key rate: General analysis} 

After $M$\ rounds of the protocol, Alice and Bob will share $M$ copies
$(\rho_{AB|L^{\prime}})^{\otimes M}$. Note that, in general, Alice and Bob
do not know anything about the physical process within the UTP, i.e., they
do not know the couple $\{\mathbf{T},L\rightarrow L^{\prime}\}$. For this
reason, what they actually get are $M$\ copies of an unknown state $\rho
_{AB}^{?}$ plus classical information $L^{\prime}$. However, by measuring a
suitable number $M^{\prime}$ of these copies, they are able to deduce the
explicit form of the conditional state $\rho_{AB|L^{\prime}}$ for the
remaining $N=M-M^{\prime}$ copies (here $M,$ $M^{\prime}$ and $N$ are large
numbers). Then, by applying local measurements, Alice on her private
systems and Bob on his, they are able to extract two correlated classical
variables, $X$ and $Y$. Finally, from these variables, they can derive a
shared secret key via the classical techniques of error correction
(EC) and privacy amplification (PA). These procedures can be implemented
using one-way classical communications between these two parties.

\begin{figure}[ptbh]
\vspace{0.2cm}
\par
\begin{center}
\includegraphics[width=0.40\textwidth]{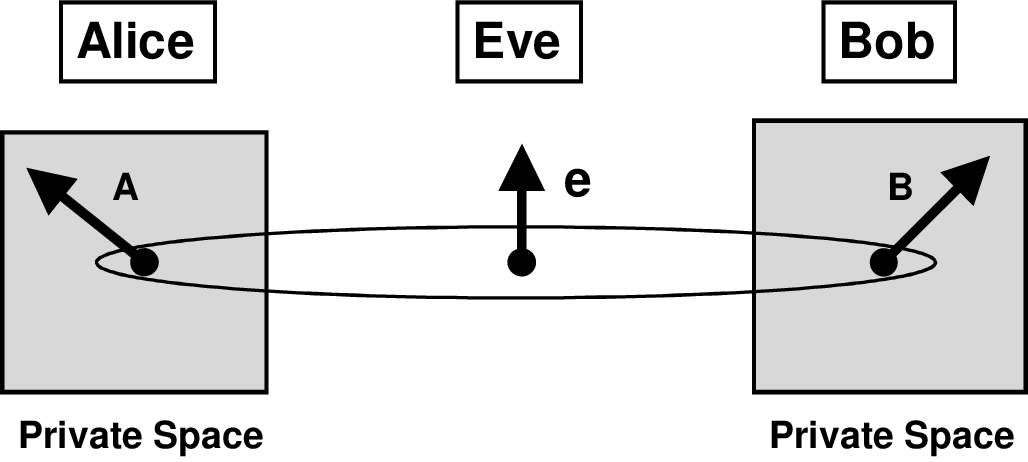}
\end{center}
\par
\vspace{-0.2cm}\caption{Purified scenario.}%
\label{ABE}%
\end{figure}

Let us bound the secret-key rate of the protocol. For simplicity we omit here
the conditioning on $L^{\prime}$, so that Eq.~(\ref{ABstate}) simply becomes
$\rho_{AB}=\mathrm{Tr}_{E}\left(  \rho_{ABE}\right)  $. It is understood that
the final result must be averaged over $L^{\prime}$. Independently from its
generation, the (generally) mixed state $\rho_{AB}$ can be purified in a pure
state $\Phi_{ABe}=\left\vert \Phi\right\rangle \left\langle \Phi\right\vert
_{ABe}$ by introducing a suitable system \textquotedblleft$e$%
\textquotedblright\ to be assigned to Eve (this is generally larger than the
$E$ system considered before). After this purification, the scenario is the
one depicted in Fig.~\ref{ABE}. Here, for every bipartition of the systems,
$\{AB,e\},$ $\{Ae,B\}$, or $\{Be,A\}$, the corresponding reduced states have
the same von Neumann entropy. In particular, we have
$S(\rho_{AB})=S(\rho_{e})$.

Now suppose that Alice performs a POVM $\mathcal{M}_{A}=\{\hat{A}(x)\}$ on her
system $A$ with classical outcome $x$. This measurement projects $\Phi_{ABe}$
onto the conditional state%
\begin{equation}
\Phi_{Be}(x)=\frac{1}{p(x)}\mathrm{Tr}_{A}\left[  \hat{A}(x)\Phi_{ABe}\hat
{A}(x)^{\dagger}\right], \label{FIBe}%
\end{equation}
where
\begin{equation}
p(x)=\mathrm{Tr}_{ABe}
\left[  \hat{A}(x)\Phi_{ABe}\hat{A}(x)^{\dagger}\right].
\end{equation}
Thus Alice encodes the stochastic variable $X=\{x,p(x)\}$\ in the
nonlocal ensemble $\mathcal{E}_{Be}\equiv\{\Phi_{Be}(x),p(x)\}$.
Given the conditional state $\Phi_{Be|X}$ of Eq.~(\ref{FIBe}), Bob
and Eve can only access their local states, respectively given by
\begin{equation}
\rho_{B}(x)=\mathrm{Tr}_{e}\left[  \Phi_{Be}(x)\right] ,
\qquad
\rho_{e}(x)=\mathrm{Tr}_{B}\left[  \Phi_{Be}(x)\right] .
\end{equation}
Thus, on his side, Bob has the ensemble $\mathcal{E}_{B}\equiv\{\rho
_{B}(x),p(x)\}$, whose measurement estimates Alice's variable $X$. Assuming
that Bob has a quantum memory, he can collect all the private systems $B$
associated to the $N$ rounds of the protocol. Then, asymptotically for
$N\rightarrow \infty$, Bob can reach the Holevo bound \cite{Holevo}
\begin{equation}
I(X:B)=S(\rho_{B})-\sum_{x}p(x)S[\rho_{B}(x)].
\end{equation}
At the same time, Eve's information is bounded by
\begin{equation}
I(X:e)=S(\rho_{e})-\sum_{x}p(x)S[\rho_{e}(x)].
\end{equation}
Assuming one-way CCs from Alice to Bob (for implementing EC and PA), we can
write the secret-key rate as a difference of Holevo informations
\cite{DW}, i.e.,%
\begin{equation}
R=I(X:B)-I(X:e).
\end{equation}
If we now assume that Alice's POVM\ is rank one, then the conditional state
$\Phi_{Be|X}$ is pure and, therefore, $\rho_{B|X}$ and $\rho_{e|X}$ have the
same entropy, i.e., $S[\rho_{B}(x)]=S[\rho_{e}(x)]$. As a consequence, we can
write%
\begin{equation}
R=S(\rho_{B})-S(\rho_{e})=S(\rho_{B})-S(\rho_{AB})=I(A\rangle B),
\label{rate1}%
\end{equation}
where $I(A\rangle B)$ is the coherent information between Alice and Bob. Thus
the secret-key rate is lower-bounded by the entanglement-distillation rate.

\vskip 0.1truein
\noindent
{\bf Secret-key rate: Detailed analysis}

Here we make a more detailed analysis which is more closely connected
to the scenario of Fig.~\ref{GEN}. In fact, the rate $R$ of
Eq.~(\ref{rate1}) comes from the general configuration of
Fig.~\ref{ABE}, which is independent from the actual process generating
the final state of Alice and Bob. If we explicitly consider the
peculiarities of the scheme of Fig.~\ref{GEN}, then we could achieve a
larger rate $R^{\ast}\geq R$. This new rate can be achieved if Alice
and Bob have some knowledge of the classical unreliability of the
UTP, i.e., of the amount of information which is ``absorbed'' by the
classical channel $L\rightarrow L^{\prime}$. Thus, if Eve tries to
tamper with the overall security by employing fake CCs, then Alice
and Bob can potentially extract a secret-key with rate larger
than the entanglement-distillation rate.

In this section, we take the different conditionings (by $L$ and
$L^{\prime}$) explicitly into account. After the CC of
$L^{\prime}=\{l^{\prime},p(l^{\prime})\}$, Alice and Bob possess
the conditional state $\rho_{AB}(l^{\prime})$ of Eq.~(\ref{ABstate}).
Let us assume that Alice performs a POVM
$\mathcal{M}_{A}=\{\hat{A}(x)\}$ on her system $A$ with classical
outcome $x$. This generates the doubly-conditional state
\begin{equation}
\rho_{B}(x,l^{\prime})
=\frac{1}{p(x|l^{\prime})}\mathrm{Tr}_{A}
\left[\hat{A} (x)\rho_{AB}(l^{\prime})\hat{A}(x)^{\dagger}\right],
\end{equation}
where
\begin{equation}
p(x|l^{\prime})= \mathrm{Tr}_{AB}\left[\hat{A}(x)\rho_{AB}(l^{\prime})\hat
{A}(x)^{\dagger}\right].
\end{equation}
Averaging over the CCs, the output of Alice's measurement is the
unconditional variable $X=\{x,p(x)\}$, where
\begin{equation}
p(x)=\sum_{l^{\prime}}p(x|l^{\prime})p(l^{\prime})
=\mathrm{Tr}_{A}\left[ \hat{A}(x)\rho_{A}\hat{A}(x)^{\dagger}\right] .
\end{equation}
This is the secret variable to be estimated by Bob. In his private
system $B$, Bob has the ensemble
\begin{equation}
\mathcal{E}_{B}=\{p(x,l^{\prime}),\rho_{B}(x,l^{\prime})\},
\end{equation}
where $p(x,l^{\prime})=p(x|l^{\prime})p(l^{\prime})$. Clearly, this
ensemble depends on both $X$\ and $L^{\prime}$. Exploiting his
knowledge of $L^{\prime }$, Bob applies a conditional measurement
$\mathcal{M}_{B|L^{\prime}}$ to his system $B$ which estimates the
value $x$ encoded by Alice. Asymptotically (i.e., for
$N\rightarrow\infty$), using a quantum memory and averaging over the
CCs (i.e., over $L^{\prime}$), Bob can reach the conditional Holevo
information \cite{EHSBOB}
\begin{equation}
I(X:B|L^{\prime})=\sum_{l^{\prime}}p(l^{\prime})\,I(X:B|L^{\prime}
=l^{\prime }). \label{HolB}
\end{equation}
For Eve we have to consider the different conditioning given by $L$.
Thus, the conditional state that Eve shares with Alice is
\begin{equation}
\rho_{AE|L}=\mathrm{Tr}_{B}\left(  \rho_{ABE|L}\right) ,
\end{equation}
which becomes $\rho_{E|XL}$ after Alice's projection. Explicitly this
state is given by
\begin{equation}
\rho_{E}(x,l)=\frac{1}{p(x|l)}\mathrm{Tr}_{A}
\left[  \hat{A}(x)\rho_{AE}(l)\hat {A}(x)^{\dagger}\right] ,
\end{equation}
where
\begin{equation}
p(x|l)=\mathrm{Tr}_{AB}
\left[\hat{A}(x)\rho_{AE}(l)\hat{A}(x)^{\dagger}\right] .
\end{equation}
Thus, Eve has the ensemble
\begin{equation}
\mathcal{E}_{E}=\{p(x,l),\rho_{E}(x,l)\},
\end{equation}
where $p(x,l)=p(x|l)p(l)$. Asymptotically, Eve can eavesdrop
$I(X:E|L)$ bits per copy \cite{EHSEVE}.

\begin{figure}[ptbh]
\vspace{-0.1cm}
\par
\begin{center}
\includegraphics[width=0.42\textwidth]{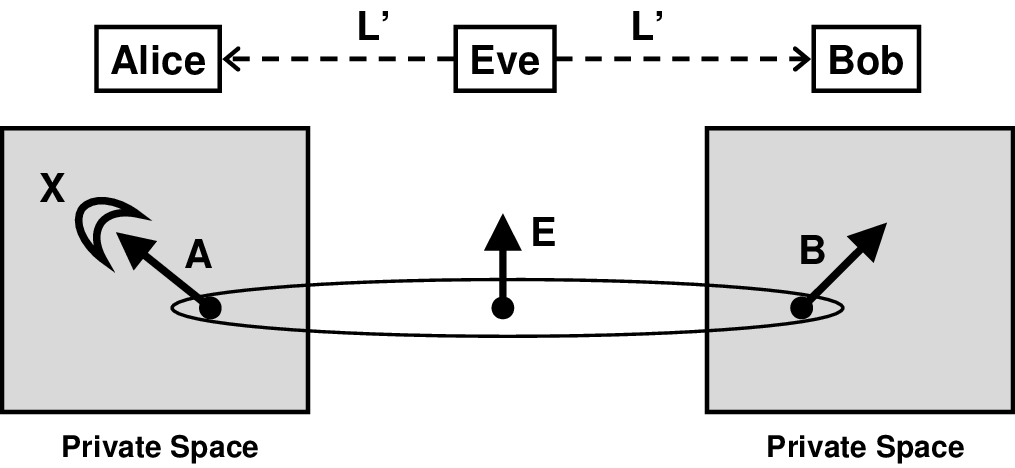}
\end{center}
\par
\vspace{-0.3cm}\caption{Conditional state $\rho_{ABE|L^{\prime}}$ projected
onto $\rho_{BE|XL^{\prime}}$.}%
\label{PicLP1}%
\end{figure}

As a result, we can write the secret-key rate
\begin{equation}
R^{\ast}=I(X:B|L^{\prime})-I(X:E|L).
\end{equation}
This quantity can be rewritten as $R^{\ast}=R^{\prime}+\Delta$,
where
\begin{equation}
R^{\prime}\equiv I(X:B|L^{\prime})-I(X:E|L^{\prime}), \label{Rprime}
\end{equation}
and
$\Delta\equiv I(X:E|L^{\prime})-I(X:E|L)$, quantifies the information
which is ``absorbed'' by the classical channel
$L\rightarrow L^{\prime}$. We call $\Delta$ the ``classical cheating''
by Eve. Clearly, we have $\Delta=0$ for $L^{\prime}=L$. $R^{\prime}$
is the ``apparent rate'', which refers to the apparent scenario where
Alice, Bob and Eve are all subject to the same conditioning
$L^{\prime}$. In other words, $R^{\prime}$ is computed assuming the
total state $\rho_{ABE|L^{\prime}}$, which is then projected onto
$\rho_{BE|XL^{\prime}}$ by Alice's measurement (see Fig.~\ref{PicLP1}).

We can now easily prove that the secret-key rate is larger than
the entanglement-distillation rate. We have the following result
(see Supplementary Material for the proof).

\textbf{Theorem}.~\textit{Suppose that Eve measures the incoming systems
but cheats on the results using a classical channel}
$L\rightarrow L^{\prime}.$ \textit{Then, Alice and Bob's secret-key rate
satisfies}
\begin{equation}
R^{\ast}\geq I(A\rangle B|L^{\prime})+\Delta,
\end{equation}
\textit{where} $I(A\rangle B|L^{\prime})$ \textit{is the coherent
information conditioned to Eve's fake variable}
$L^{\prime}$\textit{, and }$\Delta$ \textit{is the classical cheating}.

Our analysis leaves an intriguing open question. It would be wonderful
to provide an explicit example where simultaneously $\Delta>0$ and
$I(A\rangle B|L^{\prime})=0$, so that $R^{\ast}>0$. This would imply
secret-key distillation without entanglement distillation. More
generally, we cannot exclude the possibility that
$R^{\ast}>I(A\rangle B|L^{\prime})$ by using POVMs which are not rank one.


\vskip 0.1 truein \noindent {\bf Conclusion}

We have shown that virtual channels may replace real channels in
the QKD setting so as to remove any possibility of side-channel
attacks. In its simplest setting, our QKD protocol corresponds to
an entanglement swapping experiment, where the dual teleportation
channels act as ideal Hilbert space filters to wipe out
side-channel attacks. The authenticated users' private spaces are
designed so that any incoming quantum signal is topologically
excluded from access to detectors, detector settings or
state-generation settings, thus side-channel probing attacks of
the private spaces are eliminated. Finally, an external untrusted
party performs a suitable LOCC (such as a Bell-state measurement)
to create correlations necessary for shared key generation.

\vskip 0.1 truein \noindent {\bf Acknowledgments}

The research leading to these results has received funding from
EPSRC under grant No. EP/J00796X/1 (HIPERCOM).

\bigskip

\begin{center}
{\LARGE Supplementary Material}{\Large  }

\end{center}


\section{In defense of private spaces}

In quantum cryptography unconditional security proofs are derived
under the assumption that Alice's and Bob's apparata (private
spaces) are completely inaccessible by an eavesdropper who,
therefore, can only attack the signal systems which are
transmitted through the quantum communication channel connecting
the two parties. Under this assumption, secret-key rates and
security thresholds are derived in both discrete and continuous
variable quantum key distribution.

One potential loophole in the security proofs is related to how a
theoretical protocol is actually implemented experimentally. Any
redundant information encoded in extra degrees of freedom or extra
Hilbert space dimensions outside the theoretical prescription can
allow for so-called side-channel attacks. By their nature, such
attacks may be of classical or quantum degrees of freedom and are
insidious because even quantifying their threat appears to involve
understanding what have been called unknown unknowns about the
vulnerability of the experimental set-up.

Progress has been made on eliminating side-channel attacks in the
quantum communication channels between private spaces, but this
leaves open potential attacks on the private spaces through their
quantum communication ports. Let us therefore take a step back and
consider private spaces in more detail: What goes on in Alice's
and Bob's private spaces involves a significant amount of
classical information processing; at the very least the key itself
will be generated and stored as classical information. Now with
virtually any technology we have today classical information is
stored, processed and transmitted in a highly redundant fashion
(many electrons are used to charge a capacitor to represent a bit
value, or many electrons must pass through the base junction of a
transistor to effect a logical switching operation, tapping on a
keyboard produces sound waves and electromagnetic signals in
addition to the `legitimate' electrical signals in the wires,
etc). In principle any of this redundant information may leak out
of the private space through a \textquotedblleft
parasite\textquotedblright\ channel. An eavesdropper might
therefore ignore the quantum communication channel and directly
attack Alice's and Bob's apparata by exploiting the presence of
parasite channels: this is also a \textquotedblleft side-channel
attack\textquotedblright.

The implicit assumption in quantum cryptography is that we could
always improve technology in such a way that Alice's and Bob's
private spaces are not affected by the presence of parasite
channels, so that the legitimate participants do indeed have
access to absolutely private spaces. (For instance, Alice and Bob
could simulate the classical information processing on a quantum
computer. A hacked operating system on such a machine could be
tested for by randomly running subroutines that confirm that
coherence is preserved and that no information is copied out to
where it can be stored or transmitted by a trojan program --- see
also Ref.~[\onlinecite{blind}].)

However, even if you rely on a perfect isolation technology, there
remains a potential chink in this armor, which is the quantum
communication port used either to transmit a quantum state out of
your private space or to accept a quantum state for detection into
it.

If you open a communication port for quantum states to enter or
leave you must explicitly deal with side channels which can be
probing these links to your private space. Eve can potentially
send trojan systems through Alice's and Bob's communication ports
and detect their reflection to infer both state preparation and
measurement settings. As an example, in the standard BB84
protocol, Eve can irradiate Alice's apparatus by using optical
modes at slightly different frequencies. Then, from reflection,
Eve can infer the polarization chosen in each round of the
protocol. Thanks to this information, Eve can measure each signal
system in the correct basis. Another example regards the so-called
plug-and-play systems, where trojan systems can be reflected
together with signal systems, as discussed in Ref.~\cite{Gis02}.

Our paper shows how to overcome the problem of the open quantum
communication ports, therefore making feasible the notion of
absolutely private spaces. Note that this problem is not addressed
by current device-independent quantum cryptography, where such
attacks on the private space ports are simply considered
illegitimate as they violate the strong private space assumption.
The key point of our scheme is that detectors are no longer
\textquotedblleft in line\textquotedblright\ with the quantum
communication port of the private space. For this reason, it is
not possible for an external party to probe the port and obtain
detector settings or readouts from the processing of parasite
systems. In order to explain this key feature in detail, we
analyze the problem of the quantum communication ports by
comparing standard protocols with our scheme.

\begin{figure}[tpbh]
\vspace{-0.3cm}
\par
\begin{center}
\includegraphics[width=0.5\textwidth]{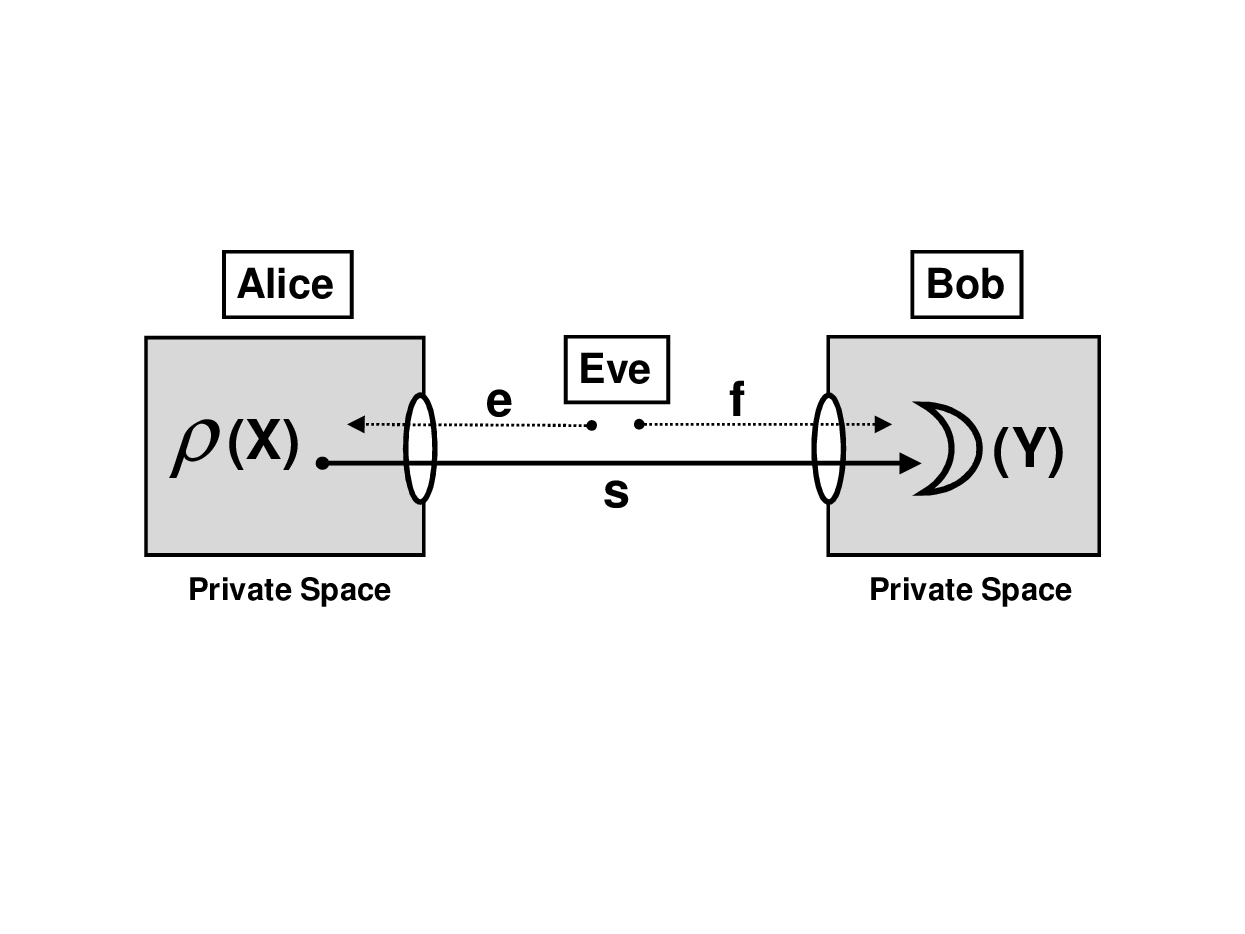}
\end{center}
\par
\vspace{-2.3cm}
\caption{Port attack in a prepare and measure protocol.}%
\label{ex1}%
\end{figure}

In Fig.~\ref{ex1}, we depict a general prepare-and-measure
protocol, where Alice's variable $X$ is encoded in a quantum state
$\rho(X)$ by modulation. Bob's variable $Y$ is the output of a
quantum measurement. Here, Eve can attack the quantum
communication ports by using two trojan systems $e$ and $f$. By
means of $e$, Eve can retrieve information about the state
preparation $X\rightarrow\rho(X)$. By means of $f$, she can
retrieve information about the measurement apparatus of Bob and,
therefore, about $Y$.

\begin{figure}[ptbh]
\vspace{-0.3cm}
\par
\begin{center}
\includegraphics[width=0.5\textwidth]{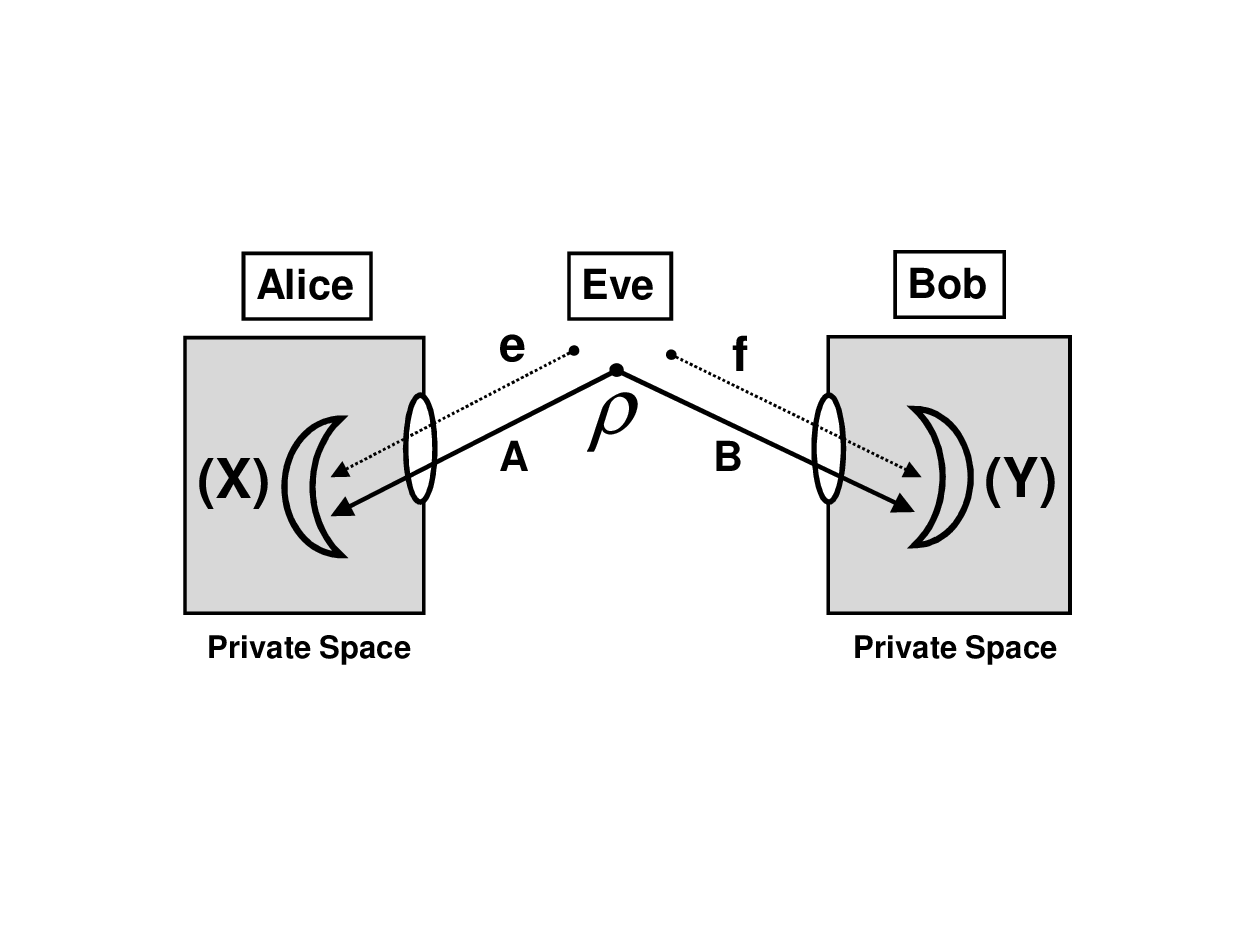}
\end{center}
\par
\vspace{-1.6cm}
\caption{Port attack in an entanglement-based protocol.}%
\label{ex2}%
\end{figure}

In Fig.~\ref{ex2}, we depict a general entanglement-based
protocol, where an untrusted party (Eve) distributes entanglement
between two parties. This is done by distributing an entangled
state $\rho=\rho_{AB}$, where system $A$ is sent to Alice and
system $B$ is sent to Bob. Alice and Bob can perform entanglement
distillation and measure the output distilled systems to derive
two correlated classical variables, $X$ and $Y$, respectively. In
this scenario, Eve can decide not to attack the source $\rho$ but
directly the two quantum communication ports of Alice and Bob. Eve
can probe these ports by using two trojan systems $e$ and $f$,
which can retrieve information about Alice's and Bob's distilling
and detecting apparata. As a result, Eve can infer information
about $X$ and $Y$.

\begin{figure}[ptbh]
\vspace{-0.3cm}
\par
\begin{center}
\includegraphics[width=0.48\textwidth]{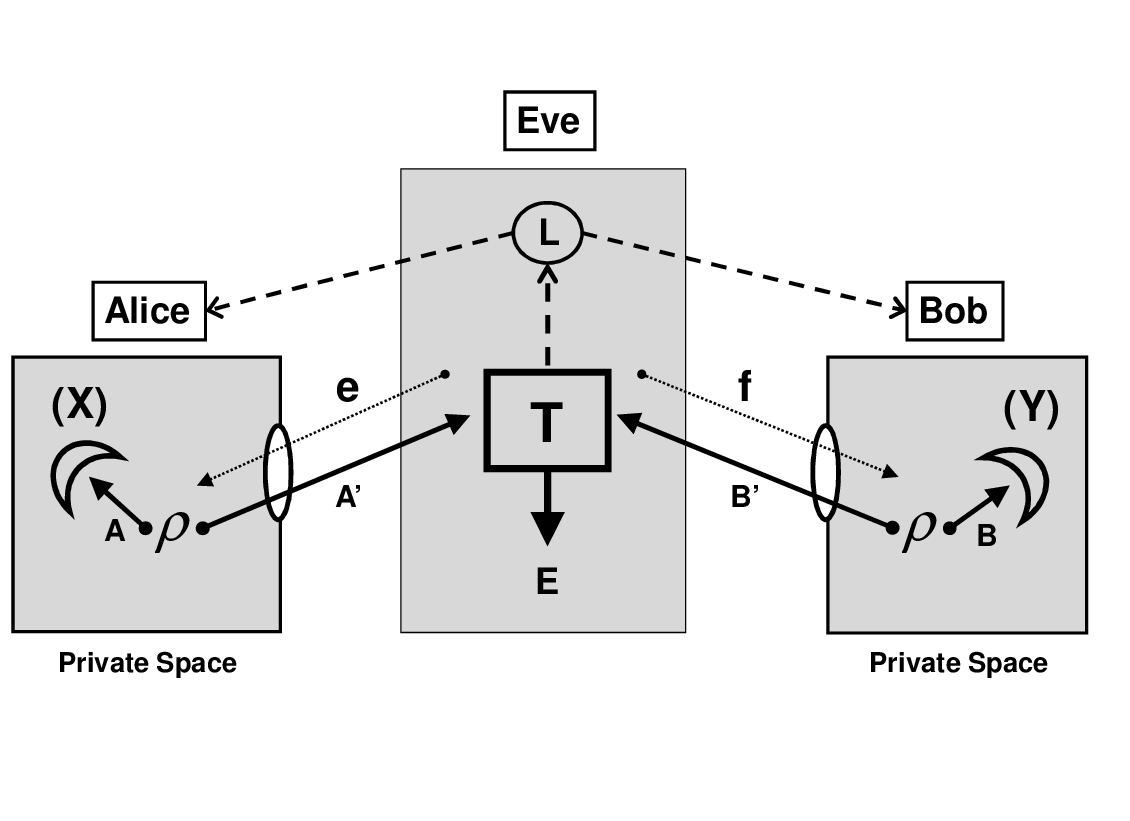}
\end{center}
\par
\vspace{-1.6cm}
\caption{Port attack in our scheme.}%
\label{ex3}%
\end{figure}

In Fig.~\ref{ex3}, we depict our protocol where an untrusted party
(Eve) represents an entanglement swapper between Alice and Bob.
This is generally done by measuring two \textit{public} systems,
$A^{\prime}$ and $B^{\prime}$, received from Alice and Bob,
processing the outcome of the measurement, and classically
communicating the processed data back to Alice and Bob. As a
result the two private systems, $A$ and $B$, become correlated, so
that Alice and\ Bob can extract two correlated classical
variables, $X$ and $Y$, by applying suitable measurements. In
particular, if Alice and Bob can access quantum memories, then
they can extract a secret key at a rate which is at least equal to
the coherent information between $A$ and $B$. Eve can attempt a
side-channel attack against the two ports by sending two trojan
systems $e$ and $f$. In this case, however, the apparata which
detect the two private systems $A$ and $B$ are inaccessible to
Eve. By exploiting reflections from the ports, Eve can only
retrieve information regarding the reduced states
$\rho_{A^{\prime}}$ and $\rho_{B^{\prime}}$ of the two public
systems $A^{\prime}$ and $B^{\prime}$. However, these reduced
states contain no useful information about the private system $A$
or $B$ or Alice's or Bob's detector settings or outputs.

\begin{figure}[ptbh]
\vspace{-0.8cm}
\par
\begin{center}
\includegraphics[width=0.45\textwidth]{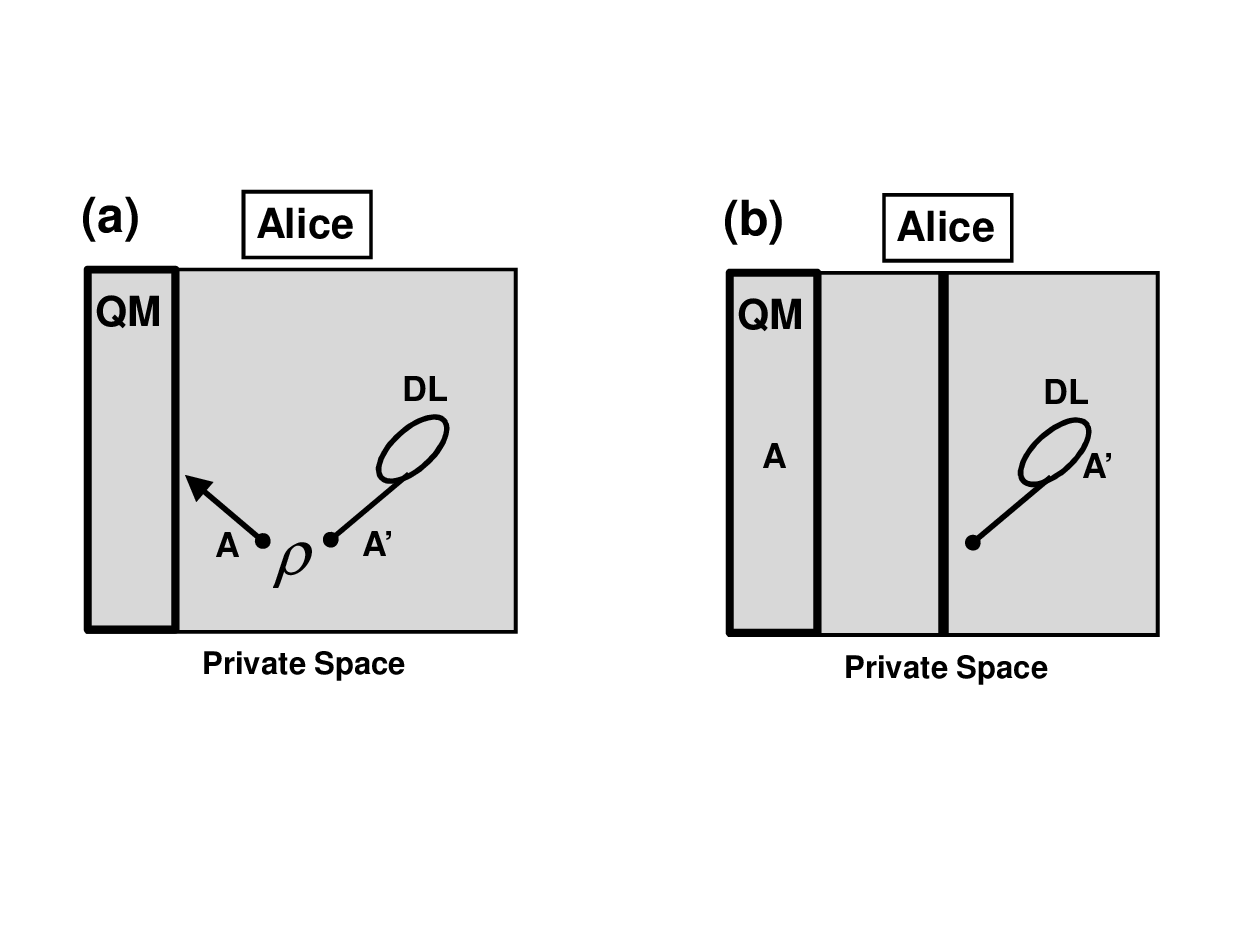}
\end{center}
\par
\vspace{-1.3cm}
\par
\begin{center}
\includegraphics[width=0.45\textwidth]{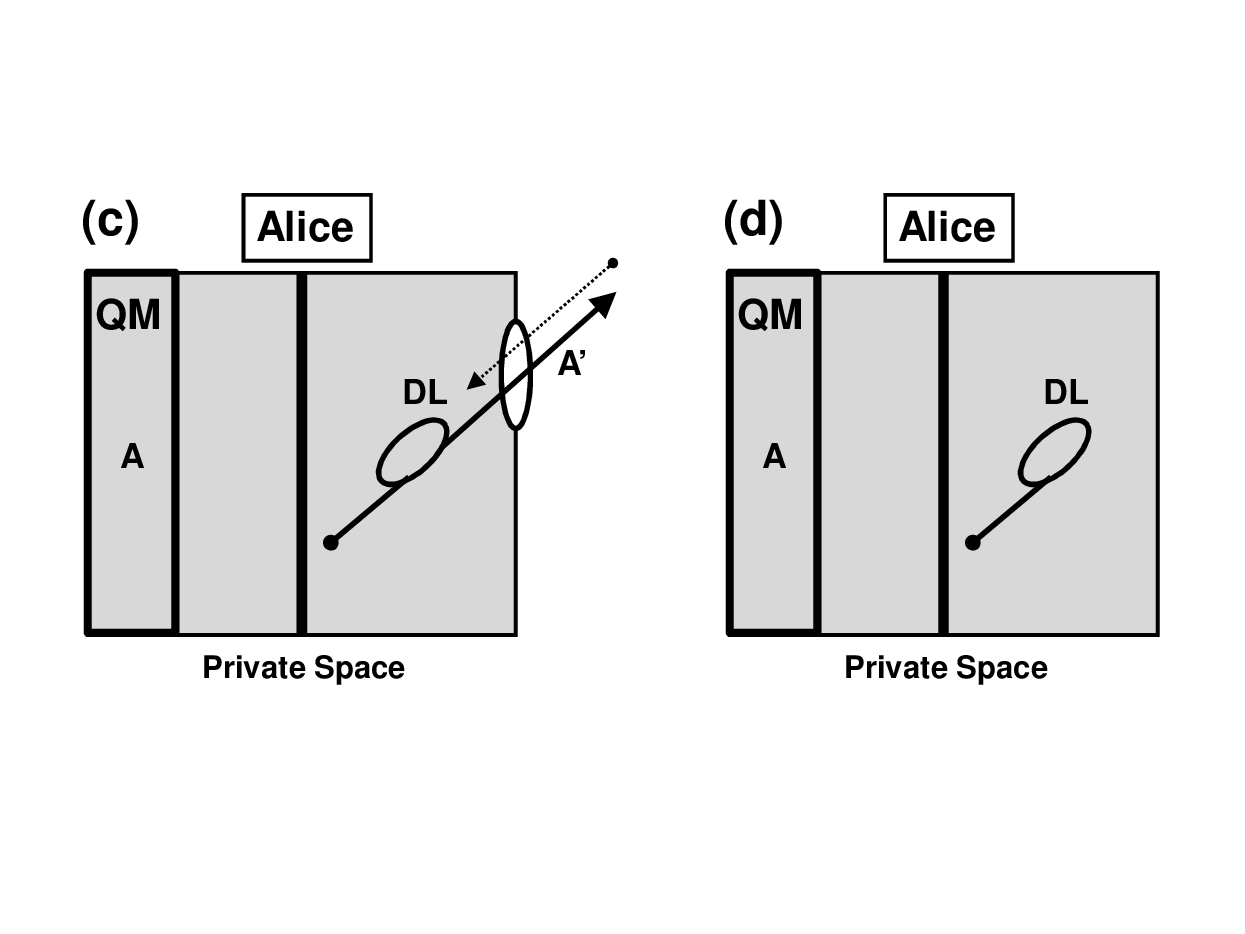}
\end{center}
\par
\vspace{-1.3cm}
\par
\begin{center}
\includegraphics[width=0.45\textwidth]{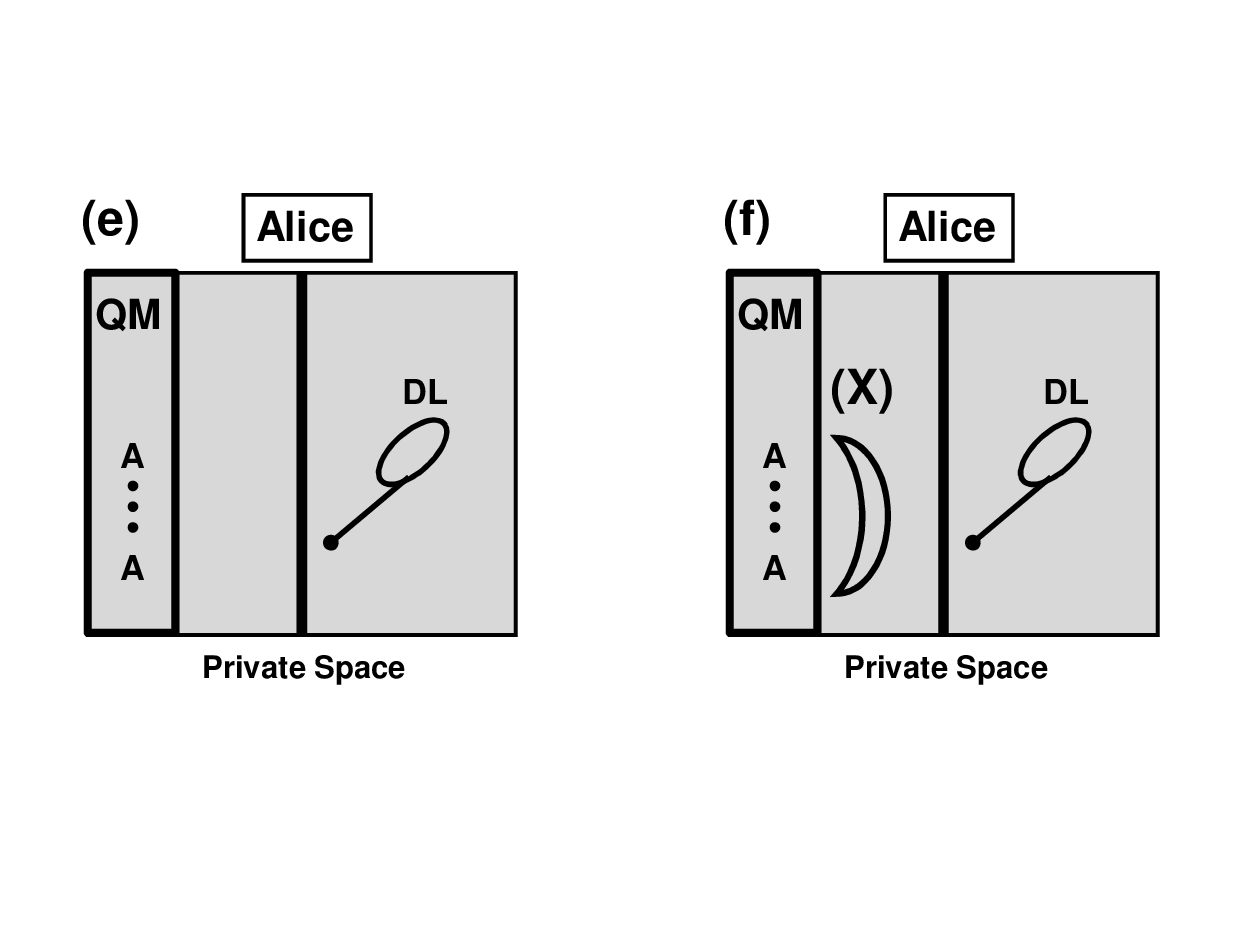}
\end{center}
\par
\vspace{-1.6cm} \caption{Possible procedure for the full isolation
of the private systems. See text for explanations.}%
\label{ports}%
\end{figure}

To understand better how the full isolation of the private systems
might be achieved, we may consider the procedure depicted in
Fig.~\ref{ports}. It is explained for Alice's private space, but
steps are identical for Bob.

In the first step (a), Alice's port is closed and she prepares an
entangled state $\rho=\rho_{AA^{\prime}}$ where system $A$ is
directed towards a quantum memory (QM), while system $A^{\prime}$
is directed towards a delay line (DL). In step (b), once system
$A$ is stored in the memory and while system $A^{\prime}$ is
trapped in the delay line, a shutter is used to fully separate the
delay line from the rest of Alice's apparatus. Note that a virtual
channel between $A$ and $A^{\prime}$ has been created. In step
(c), Alice's quantum communication port is opened and system
$A^{\prime}$ is transmitted to Eve. During this stage, trojan
systems may enter the port but no detector is in line with the
port. In step (d), the port is closed with the private system $A$
kept in the memory. The previous steps (a)-(d) are repeated many
times, so that Alice collects many private systems in her quantum
memory. We therefore reach step (e) of the figure. Finally, once
Alice has received all the classical communications, she applies a
collective quantum measurement on her quantum memory to retrieve
the classical variable $X$. This measurement can include or be
anticipated by an entanglement distillation.

\section{Notation and basic formulas}

In part of the derivation we adopt the enlarged Hilbert space
(EHS) representation, where stochastic classical variables are
embedded in quantum systems. Consider a stochastic variable
$X=\{x,p(x)\}$ which is encoded into an ensemble of states of some
quantum system $A$, i.e.,
\begin{equation}
\mathcal{E}_{A}=\{p(x),\rho_{A}(x)\}. \label{ensemble}%
\end{equation}
This ensemble may be equivalently represented by the
classical-quantum (CQ) state
\begin{equation}
\rho_{\mathbf{X}A}=\sum_{x}p(x)\left\vert x\right\rangle
\left\langle
x\right\vert _{\mathbf{X}}\otimes\rho_{A}(x), \label{CQ}%
\end{equation}
where the stochastic variable $X$ is embedded into the dummy
quantum system $\mathbf{X}$, by using an orthonormal basis
$\{\left\vert x\right\rangle \}$ in the Hilbert space
$\mathcal{H}_{\mathbf{X}}$ of $\mathbf{X}$. We denote by
$\rho_{A}(x)$ the state of a system $A$ which is conditioned by
the value $x$ of a stochastic variable $X$. The notation
$\rho_{A|X}$ refers to the conditional state $\rho_{A}(x)$ where
$x$ is not specified. Clearly, we have
\begin{equation}
\rho_{A}=\sum_{x}p(x)\rho_{A}(x).
\end{equation}

Given a quantum system $A$ in a state $\rho_{A}$, its von Neumann
entropy $S(\rho_{A})$ is also denoted by $H(A)$. Given a quantum
system $\mathbf{X}$, embedding the stochastic variable $X$, its
quantum entropy $H(\mathbf{X})$ is just the Shannon entropy
$H(X)$. Given two quantum systems, $A$ and $B$, we denote by
$I(A:B)$ their quantum mutual information. This is defined by
\begin{equation}
I(A:B)=H(B)-H(B|A), \label{mutual}%
\end{equation}
where
$H(B|A)=H(AB)-H(A)$
is the conditional quantum entropy. Note that $H(B|A)$ can be
negative and it is related to the coherent information by the
relation
\begin{equation}
I(A\rangle B)=-H(B|A).
\end{equation}
For $A=\mathbf{X}$, the quantum mutual information
$I(A:\mathbf{X})$, which is computed over the CQ-state of
Eq.~(\ref{CQ}), corresponds to the Holevo information $I(A:X)$,
computed over the ensemble of Eq.~(\ref{ensemble}). For
$A=\mathbf{X}$ and $B=\mathbf{Y}$, embedding two stochastic
variables $X$ and $Y$, $I(\mathbf{X}:\mathbf{Y})$ is just the
classical mutual information $I(X:Y)$. For three quantum systems
$A$, $B$, and $C$, we can consider the conditional quantum mutual
information
\begin{equation}
I(A:B|C)=H(AC)+H(BC)-H(ABC)-H(C),
\end{equation}
which is $\geq0$ as a consequence of the strong subadditivity of
the von Neumann entropy. For a classically correlated system
$C=\mathbf{X}$, we have a probabilistic average over mutual
informations, i.e.,
\begin{equation}
I(A:B|\mathbf{X})=I(A:B|X)\equiv\sum_{x}p(x)~I(A:B|X=x).
\end{equation}
List of other useful elements:

\begin{itemize}
\item Given a tripartite quantum system $ABC$, we can use the ``chain rule''
\begin{equation}
I(A:BC)=I(A:B)+I(A:C|B).
\end{equation}

\item Invariance of the Holevo information under addition of classical
channels, i.e., for a classical channel
\begin{equation}
p(y|x):X\rightarrow Y,
\end{equation}
we have
\begin{equation}
I(A:X)=I(A:XY).
\end{equation}

\item Given a Markov chain $X\rightarrow Y\rightarrow Z$, the classical mutual
information decreases under conditioning \cite{Cover}, i.e.,
\begin{equation}
I(X:Y|Z)\leq I(X:Y).
\end{equation}
Notice that, for three general stochastic variables, we have
$I(X:Y|Z)\gtreqless I(X:Y)$, so that the so-called ``interaction
information''
\begin{equation}
I(X:Y:Z)\equiv I(X:Y|Z)-I(X:Y),
\end{equation}
can be positive, negative or zero.

\item Data processing inequality. For a Markov chain $X\rightarrow
Y\rightarrow Z$, we have
\begin{equation}
H(X)\geq I(X:Y)\geq I(X:Z).
\end{equation}

\end{itemize}

\section{Proof of the theorem\label{APPtheorem}}

\begin{figure}[ptbh]
\vspace{-0.0cm}
\par
\begin{center}
\includegraphics[width=0.45\textwidth]{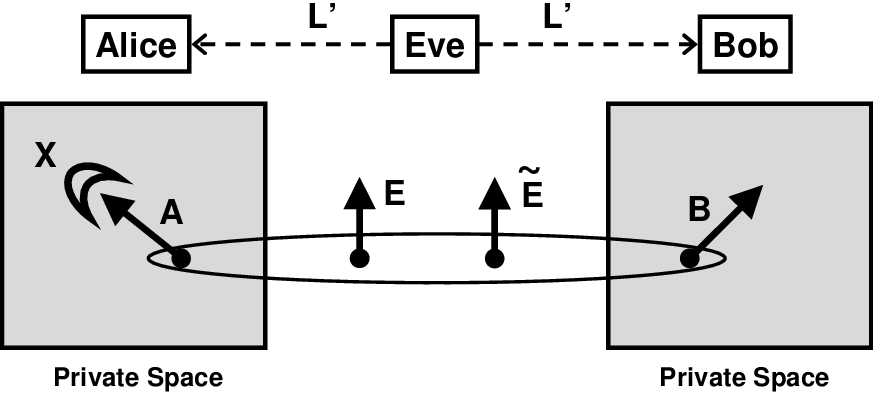}
\end{center}
\par
\vspace{-0.2cm} \caption{Purification. Conditional state
$\Phi_{ABE\tilde
{E}|L^{\prime}}$ projected onto $\Phi_{BE\tilde{E}|XL^{\prime}}$.}%
\label{PicLP2}%
\end{figure}

Let us purify the mixed state $\rho_{ABE|L^{\prime}}$ into the
pure state $\Phi_{ABE\tilde{E}|L^{\prime}}=\left\vert
\Phi\right\rangle \left\langle \Phi\right\vert
_{ABE\tilde{E}|L^{\prime}}$ by introducing an ancillary system
$\tilde{E}$ which is assumed to be in Eve's hands (so that Eve's
global system consists of $E\tilde{E}$). This scenario is depicted
in Fig.~\ref{PicLP2}.

Thus, for the total state $\rho_{ABE|L^{\prime}}$, we have
\begin{equation}
\rho_{ABE}(l^{\prime})=\mathrm{Tr}_{\tilde{E}} \left[
\Phi_{ABE\tilde{E} }(l^{\prime})\right]  .
\end{equation}
For the conditional state $\rho_{BE|XL^{\prime}}$, generated by
the measurement, we can write
\begin{align}
\rho_{BE}(x,l^{\prime})  &  =\frac{1}{p(x|l^{\prime})}
\mathrm{Tr}_{A}\left[
\hat{A}(x)\rho_{ABE}(l^{\prime})\hat{A}(x)^{\dagger}\right] \nonumber\\
&  =\frac{1}{p(x|l^{\prime})}\mathrm{Tr}_{A\tilde{E}} \left[  \hat{A}%
(x)\Phi_{ABE\tilde{E}}(l^{\prime}) \hat{A}(x)^{\dagger}\right] \nonumber\\
&  =\mathrm{Tr}_{\tilde{E}}\left[
\Phi_{BE\tilde{E}}(x,l^{\prime})\right]  ,
\label{TraceEQ}%
\end{align}
where
\begin{equation}
\Phi_{BE\tilde{E}}(x,l^{\prime})\equiv\frac{1}{p(x|l^{\prime})}\mathrm{Tr}%
_{A}\left[
\hat{A}(x)\Phi_{ABE\tilde{E}}(l^{\prime})\hat{A}(x)^{\dagger
}\right]  ,
\end{equation}
represents the conditional state $\Phi_{BE\tilde{E}|XL^{\prime}}$
which is generated by the measurement in the purified scenario.
Clearly if we discard $X$, we get the reduced state
\begin{equation}
\Phi_{BE\tilde{E}|L^{\prime}}\equiv\left\langle \Phi_{BE\tilde{E}|XL^{\prime}%
}\right\rangle _{X} =\mathrm{Tr} _{A}\left[  \Phi_{ABE\tilde{E}|L^{\prime}%
}\right]  .
\end{equation}
Because of Eq.~(\ref{TraceEQ}), the conditional state
$\Phi_{BE\tilde {E}|XL^{\prime}}$ can be used to compute
$R^{\prime}$ via
\begin{align}
R^{\prime}  &  \equiv
I(X:B|L^{\prime})_{\rho}-I(X:E|L^{\prime})_{\rho
}\nonumber\\
&  =I(X:B|L^{\prime})_{\Phi}-I(X:E|L^{\prime})_{\Phi},
\end{align}
where $\rho=\rho_{BE|XL^{\prime}}$ and
$\Phi=\Phi_{BE\tilde{E}|XL^{\prime}}$ (the computation is exactly
the same up to a trace over $\tilde{E}$). In the EHS
representation, the conditional state
$\Phi_{BE\tilde{E}|XL^{\prime}}$ becomes
\begin{equation}
\Psi_{\mathbf{XL}^{\prime}BE\tilde{E}}=\sum_{x,l^{\prime}}
p(x,l^{\prime })\left\vert x\right\rangle \left\langle
x\right\vert _{\mathbf{X}} \otimes\left\vert
l^{\prime}\right\rangle \left\langle l^{\prime}\right\vert
_{\mathbf{L}^{\prime}} \otimes\Phi_{BE\tilde{E}}(x,l^{\prime}).
\end{equation}
Thus, we can also set
\begin{equation}
R^{\prime}=I(\mathbf{X}:B|\mathbf{L}^{\prime})_{\Psi}
-I(\mathbf{X} :E|\mathbf{L}^{\prime})_{\Psi},
\end{equation}
where $\Psi=\Psi_{\mathbf{XL}^{\prime}BE\tilde{E}}$. From the
chain rule we have
\begin{align}
I(\mathbf{X} :E\tilde{E}|\mathbf{L}^{\prime})_{\Psi}  &
=I(\mathbf{X} :E|\mathbf{L}^{\prime})_{\Psi}
+I(\mathbf{X}:\tilde{E}|E\mathbf{L}^{\prime
})_{\Psi}\nonumber\\
&  =I(\mathbf{X}:E|\mathbf{L}^{\prime})_{\Psi}+\gamma,
\end{align}
where
$\gamma\equiv I(\mathbf{X}:\tilde{E}|E\mathbf{L}^{\prime})_{\Psi}
\geq 0$
\label{gamma}%
is the information contribution due to the purification
\cite{IntroEHS}. In other words, the (conditional) Holevo
information can only increase with the purification, i.e.,
\begin{equation}
I(X:E\tilde{E}|L^{\prime})=I(X:E|L^{\prime}) +\gamma\geq
I(X:E|L^{\prime}).
\end{equation}
As a consequence, we have $R^{\prime}=R^{\prime\prime}+\gamma$,
where
\begin{equation}
R^{\prime\prime}\equiv
I(X:B|L^{\prime})_{\Phi}-I(X:E\tilde{E}|L^{\prime })_{\Phi}.
\end{equation}
In terms of conditional entropies, we have
\begin{align}
R^{\prime\prime} =  &  \,H(B|L^{\prime})_{\Phi}
-H(B|XL^{\prime})_{\Phi
}\nonumber\\
&  -[H(E\tilde{E}|L^{\prime})_{\Phi}
-H(E\tilde{E}|XL^{\prime})_{\Phi}].
\label{Rsec}%
\end{align}
Here $H(E\tilde{E}|L^{\prime})$ is computed over
$\Phi=\Phi_{BE\tilde {E}|XL^{\prime}}$ discarding $X$ and $B$,
i.e., over the reduced state
\begin{equation}
\Phi_{EE|L^{\prime}}=\mathrm{Tr}_{AB} \left[  \Phi_{ABE\tilde{E}|L^{\prime}%
}\right]  .
\end{equation}
Now since $\Phi_{ABE\tilde{E}|L^{\prime}}$ is pure, we have
$H(E\tilde {E}|L^{\prime})=H(AB|L^{\prime})$, where
$H(AB|L^{\prime})$ can be computed over $\rho_{AB|L^{\prime}}
=\mathrm{Tr}_{E\tilde{E}}[\Phi_{ABE\tilde {E}|L^{\prime}}]$.
Clearly, also $H(B|L^{\prime})_{\Phi}$ can be computed over
$\rho_{AB|L^{\prime}}$. As a consequence we can recognize in
Eq.~(\ref{Rsec}) the conditional coherent information
\[
I(A\rangle B|L^{\prime})=H(B|L^{\prime})-H(AB|L^{\prime}),
\]
associated with Alice and Bob's conditional state
$\rho_{AB|L^{\prime}}$. Thus, we can set
\begin{equation}
R^{\prime\prime}=I(A\rangle B|L^{\prime})
+[H(E\tilde{E}|XL^{\prime})_{\Phi }-H(B|XL^{\prime})_{\Phi}].
\end{equation}
Here, we can assume that Alice's measurement is a rank one POVM.
As a result, $\Phi=\Phi_{BE\tilde{E}|XL^{\prime}}$ is also a pure
state, and we can set
$H(E\tilde{E}|XL^{\prime})_{\Phi}=H(B|XL^{\prime})_{\Phi}$, so
that $R^{\prime\prime}=I(A\rangle B|L^{\prime})$. Finally, we can
write
\begin{align}
R^{\ast}  &  =R^{\prime\prime}+\gamma+\Delta\nonumber\\
&  =I(A\rangle B|L^{\prime})+\gamma+\Delta\nonumber\\
&  \geq I(A\rangle B|L^{\prime})+\Delta,
\end{align}
where we have used $\gamma\geq 0$ from its definition.



\begin{thebibliography}{99}

\bibitem{BB84} C.\ H.\ Bennett, and G.\ Brassard,
in \textit{Proceedings of the IEEE International Conference on Computers,
Systems and Signal Processing}, (Bangalore, India, 1984), p.~175.

\bibitem{Ekert} A.\ Ekert, Phys.\ Rev.\ Lett.\ \textbf{67}, 661 (1991).

\bibitem{Hillery} M.\ Hillery, Phys.\ Rev.\ A \textbf{61}, 022309 (2000).

\bibitem{Cerf} N.\ J.\ Cerf, M.\ L\'evy, and G.\ Van Assche,
Phys.\ Rev.\ A \textbf{63}, 052311 (2001).

\bibitem{Grangier}
F.\ Grosshans, {\it et al.},
Nature \textbf{421}, 238 (2003).

\bibitem{Weed} A.\ M. Lance, {\it et al.},
Phys.\ Rev.\ Lett.\ \textbf{95}, 180503 (2005).

\bibitem{Scarani} V.\ Scarani, {\it et al.},
Rev.\ Mod.\ Phys.\ \textbf{81}, 1301 (2009).

\bibitem{Net} SECOQC, 2007, http://www.secoqc.net.


\bibitem{GQI} C. Weedbrook, S. Pirandola, R. G. Patron, N. J. Cerf, T. C. Ralph,
J. H. Shapiro, and S. Lloyd, Rev. Mod. Phys. {\bf 84}, 621 (2012).

\bibitem{Gis02} N.\ Gisin, {\it et al.},
Rev.\ Mod.\ Phys.\ {\bf 74}, 145 (2002).

\bibitem{Lut09} N.\ L\"utkenhaus and A.\ J.\ Shields,
New J.\ Phys.\ {\bf 11}, 045005 (2009).

\bibitem{SCA}
B.\ Qi, {\it et al.}, Quantum Inform.\ Comput.\ {\bf 7}, 73 (2007);
C.-H.\ F.\ Fung, {\it et al.}, Phys.\ Rev.\ A {\bf 75}, 032314 (2007);
Y.\ Zhao, {\it et al.}, Phys.\ Rev.\ A {\bf 78}, 042333 (2008);
L.\ Lydersen, {\it et al.}, Nature Photonics {\bf 4}, 686 (2010);
L.\ Lydersen, {\it et al.}, Nature Photonics {\bf 4}, 801 (2010);
I.\ Gerhardt, {\it et al.}, Nature Comm.\ {\bf 2}, 349 (2011); L.\
Lydersen, {\it et al.}, New J. Phys. {\bf 13}, 113042 (2011).




\bibitem{DIQKD}
D.\ Mayers and A.\ Yao, Quantum Inform.\ Comput.\ {\bf 4}, 273
(2004); J.\ Barrett, L.\ Hardy and A.\ Kent, Phys.\ Rev.\ Lett.\
{\bf 95}, 010503 (2005); A.\ Acin, {\it et al.}, Phys.\ Rev.\
Lett.\ {\bf 97}, 120405 (2006); A.\ Acin, {\it et al.}, Phys.\
Rev.\ Lett.\ {\bf 98}, 230501 (2007); N.\ Gisin, S.\ Pironio and
N.\ Sangouard, Phys.\ Rev.\ Lett.\ {\bf 105}, 070501 (2010).







\bibitem{LoChau} H.-K.\ Lo and H.\ F.\ Chau,
Science {\bf 283}, 2050 (1999).

\bibitem{Biham} E.\ Biham, B.\ Huttner and T.\ Mor,
Phys.\ Rev.\ A {\bf 54}, 2651 (1996);
H.\ Inamori, Algorithmica {\bf 34}, 340 (2002).

\bibitem{CPTP} Summing over $l$, we have a completely positive trace
preserving (CPTP) map.

\bibitem{Holevo} A.\ S.\ Holevo,
Probl.\ Inform.\ Transm.\ \textbf{9}, 177 (1973).

\bibitem{DW} I.\ Devetak and A.\ Winter,
Proc.\ R.\ Soc.\ Lond.\ A \textbf{461}, 207 (2005).

\bibitem{EHSBOB} Equivalently, we can adopt the EHS representation
(see Supplementary Material for details), where the ensemble
$\mathcal{E}_{B}$ and the stochastic variables $X$\ and
$L^{\prime}$ are described by a unique classical-quantum state
$\rho_{\mathbf{XL}^{\prime}B}=\sum_{x,l^{\prime}}p(x,l^{\prime})
\left\vert x\right\rangle\left\langle x\right\vert_{\mathbf{X}}
\otimes\left\vert l^{\prime}\right\rangle
 \left\langle l^{\prime}\right\vert _{\mathbf{L}^{\prime}}
\otimes\rho_{B}(x,l)$.
The Holevo quantity of Eq.~(\ref{HolB}) corresponds to the conditional
quantum mutual entropy $I(\mathbf{X}:B|\mathbf{L}^{\prime})$ computed
over this state.

\bibitem{EHSEVE} Equivalently, we can consider the classical-quantum
state
$\rho_{\mathbf{XL}E}=\sum_{x,l}p(x,l)\left\vert x\right\rangle
\left\langle x\right\vert _{\mathbf{X}}
\otimes\left\vert l\right\rangle \left\langle l\right\vert_{\mathbf{L}}
\otimes\rho_{E}(x,l)$,
and compute $I(\mathbf{X}:E|\mathbf{L})=I(X:E|L)$.

%

\bibitem{blind} S.\ Barz \textit{et al}.,
Science {\bf 335}, 303 (2012).

\bibitem {Cover}T.\ M.\ Cover and J.\ A.\ Thomas, (John Wiley and Sons,
Hoboken, New Jersey, 2006) p.~35.

\bibitem {IntroEHS}Note that the EHS\ representation has been mainly
introduced to give the correct interpretation to the definition of
$\gamma$, where a quantum system $E$\ conditions a classical
variable $X$ thanks to the embedding in a quantum system
$\mathbf{X}$.




\end{thebibliography}
\end{document}